\documentclass[twocolumn]{aastex62}

\usepackage{times}
\usepackage{graphicx}
\usepackage{epstopdf}
\usepackage{url}

\def\ergscm2 {erg\,s$^{-1}$cm$^{-2}$}

\def\cm2 {cm$^{-2}$}

\shorttitle{\textsc{GRBs in AGNs}}
\shortauthors{\textsc{Perna et al.}}

\begin{document}
        
\title{ELECTROMAGNETIC SIGNATURES OF
RELATIVISTIC EXPLOSIONS IN AGN DISKS}

\author{Rosalba Perna}
\affiliation{Department of Physics and Astronomy, Stony Brook University, Stony Brook, NY 11794-3800, USA}
\affiliation{Center for Computational Astrophysics, Flatiron Institute, New York, NY 10010, USA}

\author{Davide Lazzati}
\affiliation{Department of Physics, Oregon State University, 301 Weniger Hall, Corvallis, OR 97331, USA}

\author{Matteo Cantiello}
\affiliation{Center for Computational Astrophysics, Flatiron Institute, New York, NY 10010, USA}
\affiliation{Department of Astrophysical Sciences, Princeton University, Princeton, NJ 08544, USA}

\begin{abstract}
The disks of Active Galactic Nuclei (AGNs), traditionally studied as
the feeders of the supermassive black holes (SMBHs) at their centers, have
recently triggered a lot of interest also as hosts to massive stars
and hence their neutron star (NS) and black hole (BH) remnants. Migration traps and gas torques in these disks favor binary formation and 
enhance the rate of compact object mergers. In these
environments both long gamma-ray bursts (GRBs) from the death of 
 massive stars and short GRBs from NS-NS and NS-BH
mergers are expected. 
However, their properties in the unique environments of AGN disks have never been	studied. Here we show that
GRBs in AGNs can
display	unique features, owing to the unusual relative position of the shocks that characterize the burst evolution and the Thomson photosphere of the AGN disk. In dense environments, for example, the external shock develops before the internal shocks, leading to prompt emission powered by a relativistic reverse shock. The transient's time  evolution is also compressed, yielding afterglow emission that is much brighter and peaks much earlier than for GRBs in the interstellar medium. Additionally, in regions of the disk that are sufficiently dense and extended, the  light curves are dominated by diffusion, since the fireball is trapped inside the disc photosphere. These diffusion-dominated transients emerge on timescales of days in disks around SMBHs of $\sim 10^6 M_\odot$ to
 years for SMBHs of $\sim 10^8 M_\odot$.
 Finally, a large fraction of events, 
especially in AGNs with SMBHs $\lesssim 10^7 M_\odot$,  display time-variable absorption in the X-ray band.

\end{abstract}


\section{Introduction}

The accretion disks of Active Galactic Nuclei (AGNs) have had a long history of study, as they  are the engines powering the observed emission \citep{Lynden:1969}.
Since the pioneering work on the disk structure by \citet{Shakura:1973}, 
increasingly more sophisticated models have been developed to understand the detailed structure of the disk \citep[e.g.][]{Sirko2003,Thompson2005}, as well as to characterize the various regions surrounding the disk, and hence understand the physical reasons for the observational diversity of these sources.

In the last few years, the interest for AGN disks has expanded from their being sources of power for the supermassive black holes (SMBHs) at their centers,
to their being hosts to stars and hence the compact objects that they leave behind. 
Broadly speaking, stars can end up in the disks of AGNs via two mechanisms: in-situ formation resulting from disks becoming self-gravitating and unstable to fragmentation \citep[e.g.][]{Paczynski:1978,Goodman:2003,Dittmann:2020},  and capture from the nuclear star cluster surrounding the AGN  \citep[e.g.][]{Artymowicz:1993,Fabj2020}, as a result of momentum and energy loss as the stars interact with the disk. 

Massive stars leave behind neutron stars and black holes, which then interact with the dense environment of the disk, resulting in migration within the disk, and mergers either in migration traps or in regions of the disk where there is an enhanced probability to interact with compact objects outside the disk \citep[e.g.][]{Bellovary:2016,Secunda:2019,Tagawa:2020}.  The presence of compact objects, and their mergers, in AGN disks 
has become especially relevant in light of recent LIGO results: the detection of a binary BH merger with one of the BHs with mass above the pair instability range \citep{Abbott:2020prl}, and another merger with at least one of the two compact objects in the lower mass gap \citep{Abbott2020a}. Both can be explained in an AGN disk scenario via a combination of hierarchical mergers and accretion of compact objects initially formed via standard evolutionary channels \citep[e.g.][]{McKernan:2012,Bartos:2017,Stone:2017,McKernan:2018,Yang2019,Yang2020,Tagawa:2020}. 

In an AGN environment even binary BH mergers may be accompanied by an electromagnetic signature \citep{McKernan:2019}. Here we focus on the presence of massive stars in AGN disks. A fraction of these, whose inner regions are endowed with sufficient angular momentum to form an accretion disk around the BH formed from the core upon collapse, is expected to also give rise to long GRBs \citep[e.g.][]{MacFadyen:1999,Woosley:2006,Yoon:2006}, in addition to a supernova \citep{Woosley:2006R}. 
Long GRBs are predicted to occur mostly in low-metallicity environments \citep{Woosley:2006,Yoon:2006}, where stellar cores are more likely to spin rapidly at core collapse due to reduced stellar mass loss. 
This is supported by observations of the bulk population (e.g. \citealt{Graham:2013}), although some long GRBs may also occur at solar or even super-solar metallicities \citep{Levesque:2010,Schady:2015}. While the metallicity of AGN disks tends to be high at all redshifts \citep[e.g.][]{StorchiBergmann:1989,Maiolino:2019}, the evolution of stars in AGN disks might be substantially different than in standard galactic environments \citep{Cantiello:2020}, possibly resulting in a high fraction of rapidly rotating massive stars (Jermyn et al. in prep.). Additionally, as AGN disks favor mergers, NS-NS and NS-BH mergers are also expected. Both of these are believed to be progenitors of short GRBs \citep{Berger:2014R}, which has been confirmed for the case of an NS-NS merger \citep{Abbott:2017a,Abbott:2017b,Lazzati:2018}. 

With both long and short GRBs expected to occur in AGN disks, the question of their observability arises. Can the relativistic jets associated with these astrophysical transients  emerge from the dense environment of the AGN disk? Can their luminosity outshine the AGN copious emission at some wavelengths? And, if so, do their light curves  present peculiar features that make them distinguishable from the rich range of AGN variability, as well as from GRBs occurring in more typical galactic environments? 

Here we perform the first investigation of the properties of a relativistic jet evolving in the dense environment of an AGN disk. We  focus our analysis on the relevant radii which determine the main features of the observable radiation. These radii are discussed in Sec.~2, and their magnitude is computed for two models of AGN disks in Sec.~3, together with the relevant timescales and transient luminosity. We summarize our results and their implications in Sec.~4.

\begin{figure}
\includegraphics[width=1.0\linewidth]{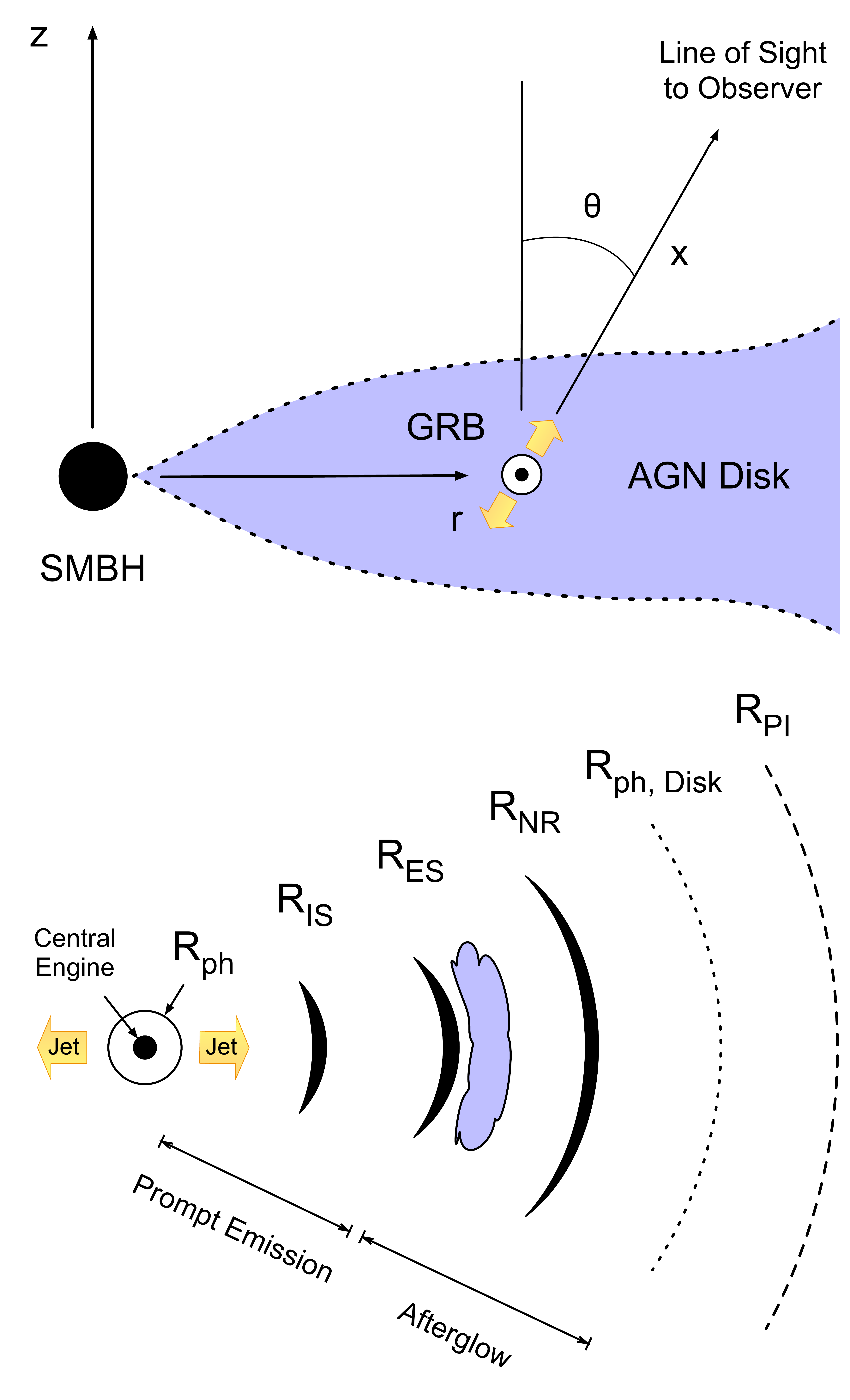}
\caption{Schematic representation of the AGN disk geometry and our coordinate system (top) and the relevant radii in commonly observed GRBs in low density environments (bottom).}
\label{fig:cartoon}
\end{figure}

\section{GRBs in AGN disks: Relevant radii and phenomenology}

The following discussion relies on the standard internal/external shock model for the production of the $\gamma$-ray prompt radiation and the following afterglow emission (e.g. \citealt{Piran1999}).  Within this scenario, the highly variable $\gamma$-ray radiation is produced via the collisions of multiple relativistic shells produced by the newly formed BH from the star's core collapse. The collision timescale between two shells is reflected in the timescale between two pulses of $\gamma$-rays. Once the bulk of the shocks have collided, the relativistic outflow begins to be decelerated by the external medium that is collecting and an external shock is formed in the ambient material, giving rise to the longer wavelength afterglow radiation. 

\begin{figure*}
\includegraphics[width=1.0\linewidth]{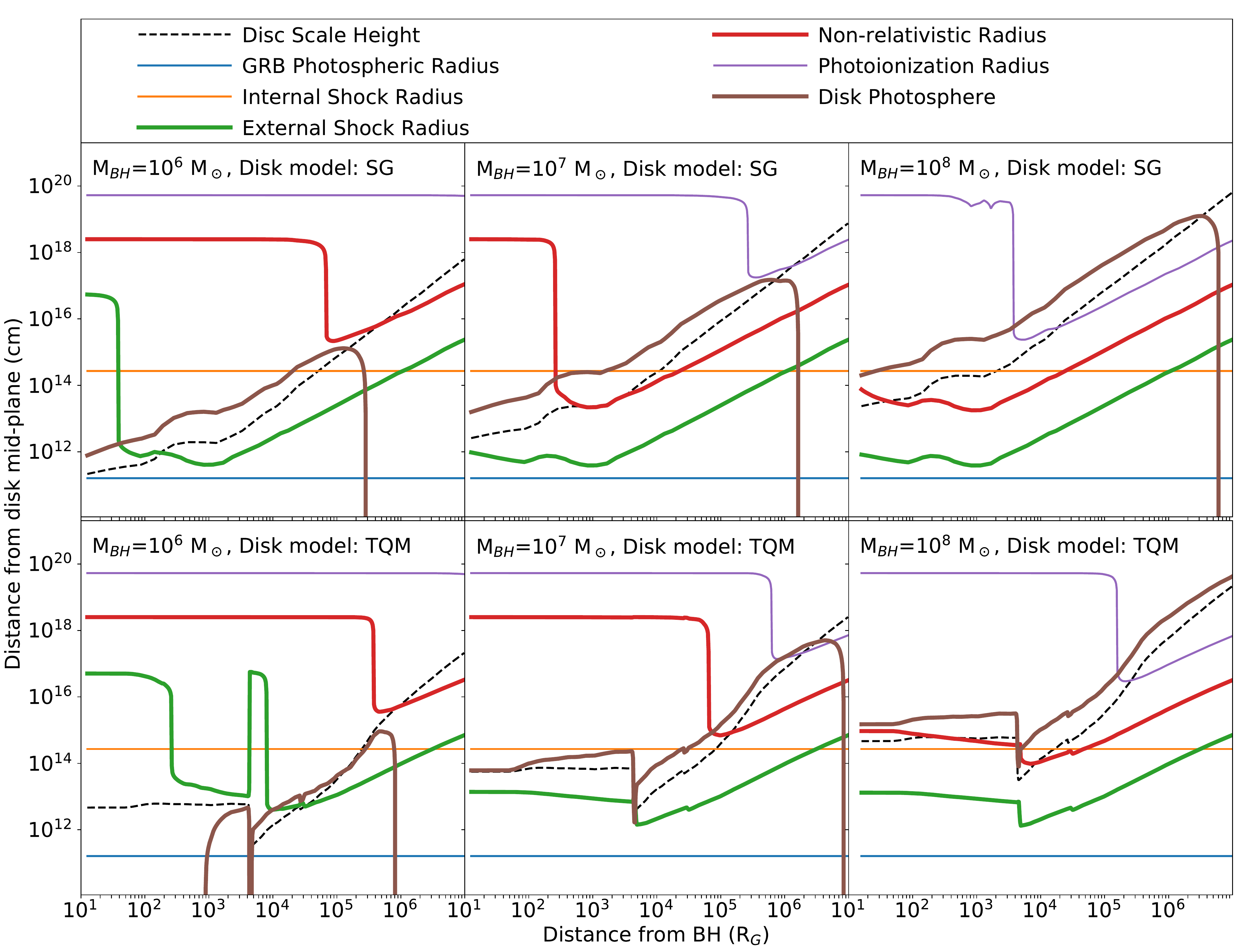}
\caption{The relevant radii for relativistic jets generated in AGN disks, for 3 values of the SMBH mass. The jet source is assumed to be located in the mid-plane of the disk, and the line of sight to the observer is perpendicular to the disk plane. The density profile of the disk is adopted from the model of \citet{Sirko2003} in the top panels and from  \citet{Thompson2005} in the bottom ones.  The disk scale height for each model is indicated with the black dashed line. }
\label{fig:Radii}
\end{figure*}

Within this model, which is schematically represented in Fig.\ref{fig:cartoon}, there are six important radii which determine the evolution of the outflow/fireball and its observability. 
These are:
\begin{itemize}
    \item The {\em photospheric radius}, i.e., the radius at which the fireball becomes transparent. This is given by \citep{Daigne2002,Lazzati2020}
\begin{equation}
    R_{\rm{ph}}=
    \frac{\sqrt{\left(\frac{cT_{\rm{eng}}}{1-\beta}\right)^2
    +\frac{3}{2\pi}\frac{L_{\rm{iso}}Y_{\rm{e}}\sigma_T T_{\rm{eng}}}{m_p c^2 \Gamma_\infty}}-
    \frac{cT_{\rm{eng}}}{1-\beta}
    }{2}\,,
    \label{eq:rph}
\end{equation}    
where $\beta$ is the outflow speed in units of the speed of light, $\Gamma_\infty$ the asymptotic Lorentz factor, $T_{\rm eng}$ is the burst engine duration, $\sigma_{\rm T}$ the Thomson cross section, $m_p$ the proton mass, $Y_e$ the electron fraction in the outflow, and
$L_{\rm{iso}}$ its isotropic luminosity. 
   
    \item The {\em internal shock radius}, i.e., the radius at which some of the bulk energy of the outflow is dissipated by collisions between shells that were ejected with different Lorentz factor. The bulk energy is transformed into internal energy that can be radiated. In the internal shock model, this is given by (e.g. \citealt{Rees1994, Piran1999})
    \begin{equation}
         R_{\rm IS}  = c\Delta t \Gamma_\infty^2\,
         \label{eq:Ris}
    \end{equation}
    where $\Delta t$ is the typical timescale between the emission of two shells, usually of the order of a fraction of a second. 
    
    \item The {\em external shock radius}, i.e., the radius at which the fireball has collected an amount of mass comparable to its rest mass divided by its Lorentz factor.
   This causes the development of a blastwave in the ambient material \citep{Meszaros1997}. The external shock radius is given by
    \begin{equation}
    R_{\rm ES} = \left(\frac{3M_{\rm fb}}{4\pi\rho\Gamma_\infty}\right)^{1/3},
\label{eq:Res}    
\end{equation}
where $\rho$ is the density of the surrounding medium and 
$M_{\rm fb} = {E_{\rm iso}}/{(c^2\Gamma_\infty)}$ 
is the fireball rest mass. The energy content of the outflow is generally parameterized in terms of its isotropic equivalent energy
$E_{\rm iso}$.
    
    \item The {\em non-relativistic radius}, i.e., the 
    radius at which the mass swiped up is large enough to cause the deceleration of the blast wave to non-relativistic speed. This radius is also known as the Sedov length and is given by (e.g., \citealt{Piran1999}):
    \begin{equation}
    R_{\rm NR}=R_{\rm ES} \Gamma_\infty^{2/3}\,.
        \label{eq:Rnr}
    \end{equation}
    
\item
The {\em disk photospheric radius} $R_{\rm ph, Disk}$, defined as the location within the disk from which the radiation can escape since the photons' mean free path for Thomson scattering becomes infinite. This is determined by solving the equation
\begin{equation}
\tau(r) = \int_{R_{\rm ph, Disk}}^\infty dx \;n_{\rm d}(x) \sigma_{\rm T}= 1\,,
    \label{eq:Rdisk}
\end{equation}
where the variable $x(r,z)$ indicates the coordinate along the line of sight to the observer, and $n_{\rm d} [x(r,z)]$ is the local number density in the disk.
    
    \item The {\em photoionization radius}, i.e., the radius out to which the ionizing radiation from the burst is abundant enough to cause complete ionization of all the elements of significant astrophysical abundance (typically, up to iron). In a cold and dense environment, this is given by:
    \begin{equation}
    R_{\rm PI} = \left[\frac{3}{4\pi n}\int_{t_0}^\infty dt
    \int_{\nu_0}^\infty d\nu \frac{ L(\nu,t)\sigma(\nu)}{h\nu}
    \right]^{1/3}\,.
        \label{Rph}
    \end{equation}
In the case of AGN disks, however, this should be considered as a lower limit, since at typical disk temperatures the gas is at least partly ionized. 
\end{itemize}

\begin{figure*}
\includegraphics[width=\linewidth]{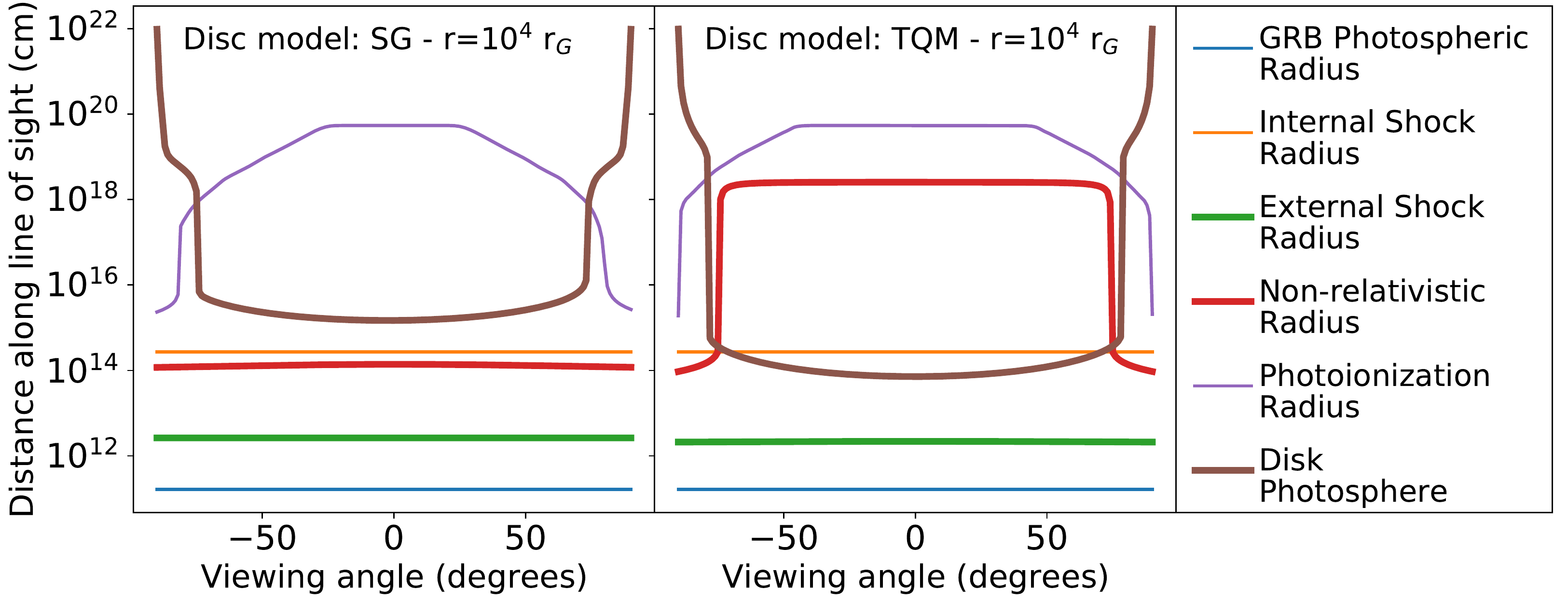}
\caption{Relevant radii as a function of the viewing angle for a representative case with SMBH mass of $10^7 M_\odot$ and a distance from the BH of $R=10^4 R_G$. }
\label{fig:view}
\end{figure*}

In the traditionally studied cases both long and short GRBs explode in interstellar medium environments, and the above list ranks the radii from the smallest to the largest. However, the ordering can change for a burst exploding in a dense and extended region, such as that of an AGN disk.
The radii affected by the ambient density are the external shock radius, the non-relativistic radius, the disk photospheric radius, and the photoionization radius. The internal shock radius and fireball photospheric radius are instead independent of the density, hence leaving the possibility that both of them, or the internal shock radius only (typically larger than the photospheric radius of the fireball) can be smaller than the external shock radius.  Under these circumstances, the developing external shock would drive a relativistic reverse shock into the jet, likely generating magnetic fields strong enough to power emission in the gamma-ray band. Note that this reverse-shock gamma-ray emission would be present irrespective of the existence of internal shocks, which are not universally accepted\footnote{In this case, the prompt emission would be released at the photosphere (e.g., \citealt{Lazzati2009}), which would be unmodified even in a disk environment.}. In the following, we assume that the radius $R_{\rm prompt}$ at which the prompt emission is generated is the smallest between $R_{\rm IS}$ and $R_{\rm ES}$. This does not consider photospheric emission, but even if that was the case, our conclusions would be unaffected. 

We can then distinguish the following situations:
\begin{itemize}
    \item $R_{\rm ph,Disk}< R_{\rm prompt}<R_{\rm ES}<R_{\rm NR}$.\\
    This is a rather 'standard' GRB, though the emission may come on faster timescales due to the higher densities (more on this in Sec.~3). The peak of the optical and X-ray afterglow emission happens at the time at which the external shock develops.
    
    \item $R_{\rm prompt}< R_{\rm ph,Disk}<R_{\rm ES}<R_{\rm NR}$.\\
In this case the prompt radiation occurs inside the disk photosphere, and hence it becomes isotropized and diluted in time via Thomson scattering.
These effects result in a reduction of the intensity by a factor $4\pi/\Omega_{\rm j}$, where $\Omega_{\rm j}$ is the angular size of the jet, and by an additional  factor $t_{\rm diff}/T_{\rm prompt}$, where  $t_{\rm diff}$ is the diffusion timescale and $T_{\rm prompt}$ is the observed duration of the prompt emission. The diffusion timescale depends on the position of the radiating source within the disk, which changes with time for a relativistic fireball. In order to keep the discussion concise, we here assume that the transients are diffused as if they were in the plane of the disk, a worst-case scenario. In this case, the diffusion timescale is given by:
\begin{equation}
    t_{\rm diff} \approx \frac{H^2 \rho_0 \sigma_T}{m_p c},
\label{eq:tdiff}
\end{equation}
where $H$ is the scale height of the disk and $\rho_0$ the density in the disk mid-plane. 
In this case the external shock develops outside of the disk photosphere, and therefore the afterglow would appear typical, though on faster timescales. As in the previous case, the luminosity peak is determined by the timescale $t_{\rm ES}$ at which the external shock is formed. 

\item $R_{\rm prompt}<R_{\rm ES}< R_{\rm ph,Disk}<R_{\rm NR}$.\\
In this case the prompt emission is diluted and scattered as in the previous case, and so too is the early afterglow radiation. 
This dilution occurs until $t_{\rm ES,d}$, at which time the external shocks exits the disk photosphere and the afterglow emission peaks. From that time on, the afterglow emission  proceeds as typical, though on faster timescales.
\item 
$R_{\rm prompt}<R_{\rm ES}< R_{\rm NR}<R_{\rm ph,Disk}$.\\
In this case both the prompt emission and afterglow emission are isotropized, time-diluted, and thermalized by Thomson scattering, both arriving to the observer on the diffusion timescale discussed above (see Eq.~\ref{eq:tdiff}).

\end{itemize}

\section{Radii in specific models of AGN disks, and observability of the transients}

In order to assess the conditions that can be verified in AGN disks, and hence the observational features that GRBs born out of AGNs would have, we consider two realistic models of AGN disks: the one by \citet{Sirko2003} (SG in the following) and the one by \citet{Thompson2005} (TQM in the following). The density profile in the disk is described by an isothermal atmosphere model,
\begin{equation}
\rho_{\rm disk} (r,z)= \rho_0(r)\exp\left[-\frac{z}{H(r)}\right]\,,
    \label{eq:rhod}
\end{equation}
with the profiles $\rho_0(r)$ (the density in the disk mid plane) and $H(r)$ (the disc scale-height)
provided by the AGN models referenced above (see Fig.1 in \citet{Fabj2020} for a visual comparison between the main properties of these disks).

\begin{figure*}
\includegraphics[width=0.95\linewidth]{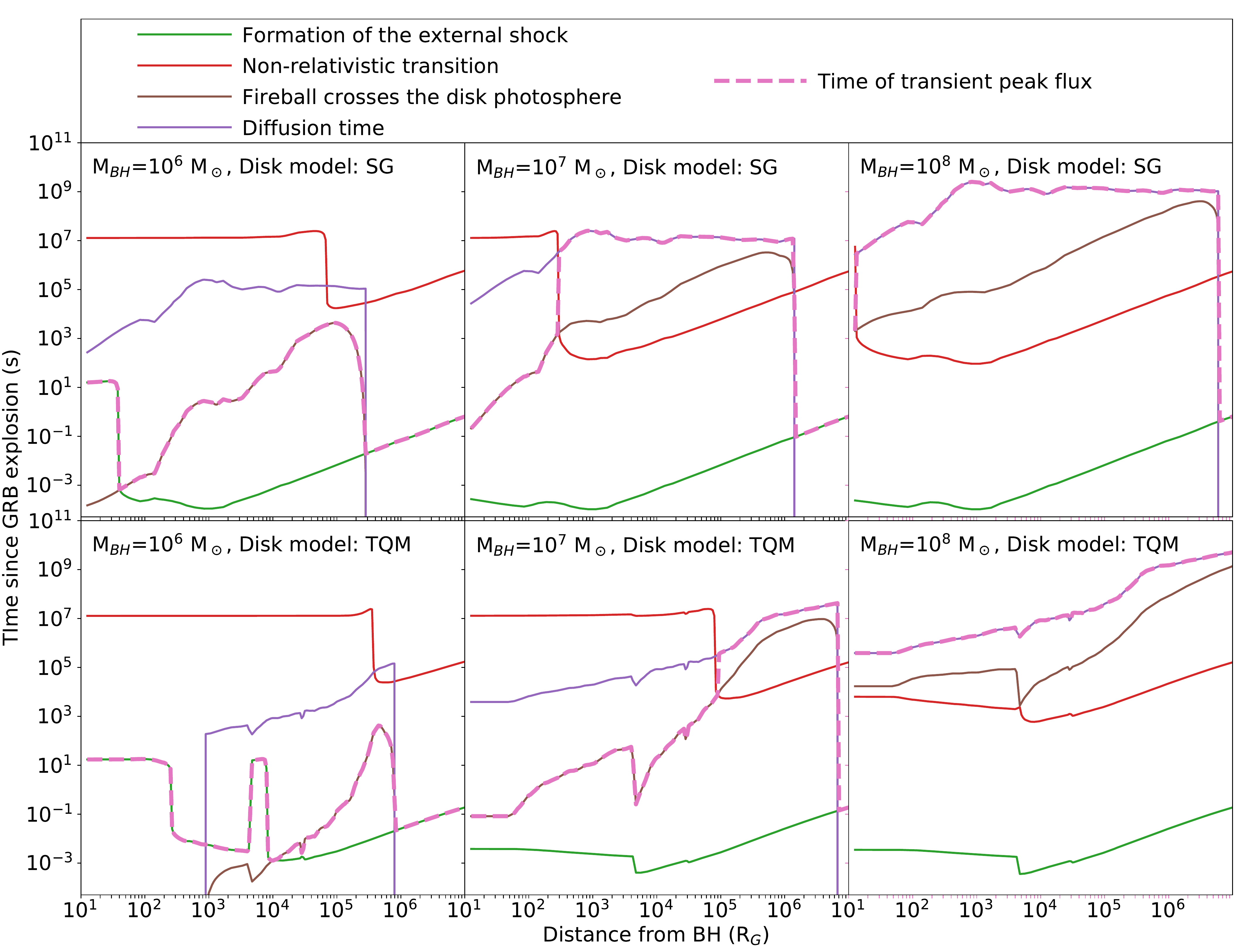}
\caption{Timescales of the afterglow transient. The observed time for the external shock formation, transition to non-relativistic expansion and crossing of the photosphere are plotted along with the diffusion timescale for a source in the disk mid plane. A thicker dashed line is used to show the expected time of maximum observed flux of the afterglow transient. The density profile of the disk is adopted from the model of \citet{Sirko2003} in the top panels and from  \citet{Thompson2005} in the bottom ones, as in Figure~\ref{fig:Radii}.}
\label{fig:aftPeak}
\end{figure*}

The relevant radii introduced in Sec.~2 are computed\footnote{We use the standard expressions for fireball evolution in the ISM. However, note that a relativistic shock  is known to accelerate in an exponential atmosphere \citep{Perna2002a} as well as in a density profile $\rho^{-k}$ with $k>4.13$ \citep{Best2000}. In an AGN disk the atmosphere is exponential in the $z$ direction, declining but shallower than  $\rho^{-4.13}$ in the outward direction, and increasing inward. A detailed radius evolution can thus only be computed numerically.} and displayed in Fig.~\ref{fig:Radii}  for the SG (top panels) and TQM (bottom panels) disk structures. In each case, we consider three values for the mass of the central SMBH: $10^6,10^7,10^8 M_\odot$, which encompass a large fraction of measured masses of SMBHs in AGNs\footnote{http://www.astro.gsu.edu/AGNmass/}. 
The black dashed line guides the eye to the location of the scale height of the disk, as compared to the relevant source-related radii. 
The location of the source is assumed to be at the disk mid-plane, while the line of sight to the observer is perpendicular to the disk plane. 
\begin{figure*}
\includegraphics[width=0.95\linewidth]{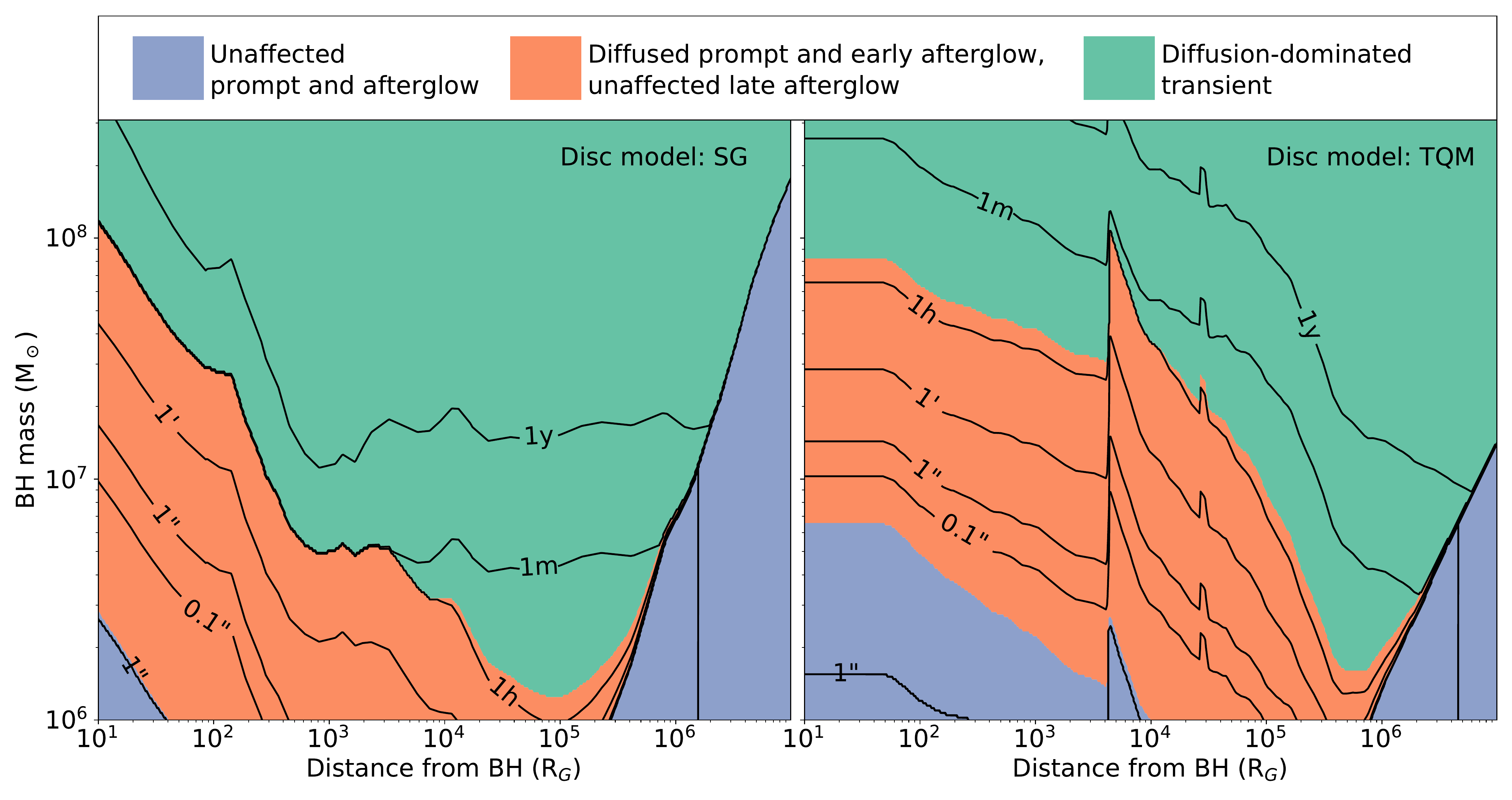}
\caption{Schematic view of the various possibilities for the outcome of a GRB from an AGN disk. {\em Blue} regions: normal GRBs. {\em Salmon}: Diffusion-dominated prompt $\gamma$-rays and early afterglows but normal late afterglow. {\em Green}: Both prompt radiation and afterglow emerge on a diffusion timescale.
The black lines indicate the time at which the electromagnetic transient peaks.}
\label{fig:2Dsummary}
\end{figure*}

For the computation of GRB-specific radii, we use
generic parameter values as typical for long GRBs (e.g. \citealt{Ghirlanda2018}):
$E_{\rm iso}=L_{\rm iso} T_{\rm eng}=10^{53}$~erg, with $T_{\rm eng}=20$~sec, $Y_e=1, \eta=300$, and $\Delta t=0.1$~sec. 
In all the circumstances considered, the photospheric radius is the smallest among all the radii. Note that this would be even more so the case for a short GRB (smaller $E_{\rm iso}$) and shorter $T_{\rm eng}$; hence the following considerations can be considered generally holding for both long and short GRBs.

Let us start by examining the $M=10^6M_\odot$ cases: in both disk models the fireball exits the disk photosphere while still relativistic, and hence there will always be some standard afterglow radiation. At small  and large disk radii
($\lesssim 10^2$ or $\gtrsim3\times10^{5}R_G$ in the SG model and $\lesssim 10^4$ or $\gtrsim10^{6}R_G$ in the TQM model, where $R_G$ is the gravitational radius), the internal and external shock radii are also outside the disk photosphere, and the full prompt and afterglow will be observed. 
A strong reverse shock is expected in most cases, when the radius of the external shock is larger than that of the internal one. The photoionization radius is well outside the disk photosphere. This implies that the early UV/X-ray flux photoionizes the medium along its line of sight, resulting in time variable absorption, if measured with time-resolved X-ray spectroscopy \citep{Perna2002b,Lazzati2002}.

As the SMBH mass increases to $10^7M_\odot$, GRBs in the outer parts of the disk ($\gtrsim 10^2\,R_G$ for the SG model and $\gtrsim 10^5\,R_G$ for the TQM disk) become non-relativistic inside the disk photosphere, and hence in those regions only longer-lived, diffusion dominated transients will be observed. At smaller radii, both the prompt emission and the early afterglow are  diluted on $t_{\rm diff}$, while the later afterglow  develops normally. Photoionization of the medium by the  GRB is also expected up to large disk radii, $\sim 10^5 R_G$.

For the case with even larger SMBH mass, $10^{8} M_\odot$, the general expectation, irrespective of the disk model, is that of a dim and long transient, diluted on the timescale $t_{\rm diff}$. 
Note that, while Fig.~\ref{fig:Radii} is calculated for a line of sight perpendicular to the disk, the outcome is similar for a large range of viewing angles $\theta$, as shown in Fig.~\ref{fig:view}. 

To further quantify the observable  properties of the transient, we compute the timescales corresponding to the relevant radii discussed above. 
These are displayed in Fig.~\ref{fig:aftPeak}. In a standard afterglow, the timescale of the peak emission at optical frequencies and above is given by that of formation of the external shock, $t_{\rm ES}$. However, if the early afterglow is  produced inside the disk photosphere  the peak emission  occurs at the time $t_{\rm ES,Disk}$, at which the fireball crosses the disk photosphere. Hence, more generally, the time of the peak emission is  the largest between $t_{\rm ES}$ (green line in Fig.\ref{fig:aftPeak}) and $t_{\rm ES, Disk}$ (brown line in the same figure). This is indicated with the dashed pink line. A remarkable feature is that the peak times are in many cases much shorter than for typical GRBs in the ISM. This is a direct result of the much higher densities in AGN disks. Peak times are longer 
when the shock becomes non-relativistic inside the disk photosphere, in which case the transient emerges on the timescale $t_{\rm diff}$, indicated with a purple line in Fig.~\ref{fig:aftPeak}. 

We summarize these results and extend them to a continuum of SMBH masses in Fig.~\ref{fig:2Dsummary}.  
 Blue areas indicate typical GRBs, for which both the prompt $\gamma$-rays and the later afterglow radiation are produced after the fireball has exited the disk photosphere.
 Salmon regions are diffusion-dominated for the prompt $\gamma$-ray transient and early afterglow, but display a normal afterglow later, when the external shock crosses the disk photosphere.
In these regions the black contours  indicate  the peak times of the afterglow: $t_{\rm ES}$ in the green regions, $t_{\rm ES, Disk}$
in the salmon regions.
 Green areas are the ones in which both the $\gamma$-ray transient and the afterglow are produced inside the disk photosphere, and hence emerge diluted on the  diffusion timescale.
The black contours in these regions indicate the diffusion time.

The extremely high densities of the medium in an AGN disk, while shortening the timescale of the emission, also make it brighter.
For an adiabatic fireball, the peak afterglow luminosity is given by \citep{Sari1998} 
\begin{equation}
L_{\nu,{\rm peak}} = 4.4\times 10^{39}{\epsilon_{B,-1}}^{1/2}E_{53}{{n}_{12}}^{1/2} {\rm erg}~{\rm s}^{-1}~{\rm Hz}^{-1}\,,
    \label{eq:Lmax}
\end{equation}
where $E_{53}=E_{\rm iso}/{10^{53}}$~erg, $n_{12}= n/10^{12}$cm$^{-3}$, and
$\epsilon_{B,-1}=\epsilon_{\rm B}/0.1$ is the fraction of shock energy that goes into magnetic energy. At an optical frequency of $\nu_O=  5\times 10^{14}$~Hz, the peak luminosity becomes $L_{\rm peak}=\nu_O L_{\nu_O,{\rm peak}}~\sim~2\times~10^{54}~ E_{53}~{{n}_{12}}^{1/2}~{\rm erg}/{\rm s}$. This maximum luminosity is  reached in the case in which the reverse shock is produced outside of the disk photosphere (with magnitude clearly dependent on the local medium density). In the case it is produced inside, the peak luminosity depends on the time at which it exits the photosphere, which in turns depends on the specific location within the disk.

To generalize the discussion of the transient luminosity while keeping it simple, we can alternatively estimate the bolometric afterglow luminosity using the observation that a fraction $\eta_{\rm aft}$ of the total fireball energy is dissipated in the afterglow. The peak luminosity can then be simply estimated as $L_{\rm peak, bol}\sim\eta_{\rm aft} E/ t_{\rm peak} = 10^{53 }E_{\rm 53} \eta_{{\rm aft}}/(t_{\rm peak}/{\rm s})$~erg~s$^{-1}$.
The brightness contrast with the AGN disk luminosity will depend on the observation band. Most AGNs have bolometric luminosities   
 $\sim10^{43 - 47}$~erg~s$^{-1}$ spread across the spectrum but with a large fraction in the UV/optical \citep[e.g.][]{Woo:2002}.   If a fraction $f_O$ of the afterglow luminosity is emitted in the optical band, then transients with $t_{\rm peak}\lesssim 2\times10^5$~s ($\sim 2$ days) will be $\sim 5$ times brighter than an AGN with an average optical luminosity of $\sim 10^{45}$~erg~s$^{-1}$. Here we have assumed $\eta_{\rm aft}f_O=10^{-2}$, an estimate that yields Eq.~\ref{eq:Lmax} for a non-diffused transient. Note, however, that for diffusion dominated transients the peak luminosity decreases by a factor $4\pi/\Omega_j\sim100$ due to the isotropization of the beamed afterglow photons. In other bands, such as X-rays, IR and radio, the transient will generally be  easily detectable against the background AGN. Multi-wavelength monitoring will be key to distinguish diffusion-dominated GRB afterglows from other transients, since their energetics and timescales would be similar to those of core-collapse supernovae, but their spectra are expected to remain much wider in frequency. Multimessenger signals, such as gravitational waves and/or neutrinos, could also be insightful for nearby events.

\section{Summary} 

AGN disks are emerging as rich environments for hosting stars and the NSs and BHs  that the most massive ones leave behind upon their deaths.  Formation of binary compact objects and mergers are enhanced in disk environments. 
 Both long and short GRBs are hence
expected to occur in AGN disks, the former from a fraction of massive and fastly rotating stars, the latter from NS-NS and possibly NS-BH mergers.

Both types of transients are  produced by relativistic fireballs, and here we have addressed the timely question of their evolution in the special environments of AGN disks, and the observability of the prompt and afterglow radiation that they produce. Our analysis has uncovered some unique features, especially evident in the lower mass SMBHs, in which cases the fireball emerges from the disk photosphere while still relativistic. In particular, due to the very high densities of AGN disks, the external shock generally forms before the internal shocks, hence potentially yielding a strong reverse shock. Under these conditions, while no
$\gamma$-rays would be formed in internal shocks, they may be produced by the powerful reverse shock.

 The time at which the afterglow emission peaks is determined, for a large fraction of disk radii and as long as the fireball remains relativistic up to the disk photosphere, by the time it takes for the fireball to cross the disk photosphere, rather than by the typical timescale for the external shock to form. 
On the other hand, in disk regions in which the fireball becomes non-relativistic prior to exiting the disk photosphere, the radiation will emerge on the diffusion timescale, which varies from between a few days to a few years for AGN disks around SMBHs of masses in the $10^6-10^8M_\odot$ range. 
Last, for transients with early monitoring of the X-ray emission, time-variable absorption is generally expected. 
Future works will be devoted to further explore and quantify via numerical simulations several aspects of this first analysis.

\acknowledgements 
RP acknowledges support by NSF awards AST-1616157 and AST-2006839
and from NASA (Fermi) award 80NSSC20K1570.
DL acknowledges support from NASA grant NNX17AK42G (ATP) and NSF grant AST-1907955.
The Center for Computational Astrophysics at the Flatiron Institute is supported by the Simons Foundation.

\bibliographystyle{aasjournal}
\bibliography{refs}

\end{document}